\documentclass[11pt]{article}
\usepackage{amsfonts,amsmath,amssymb,graphics,epsfig,subfig}

\begin{document}

\date{\empty}

\title{\textbf{Friedmann-like universes with weak torsion: a dynamical system approach}}

\author{John D. Barrow$^1$, Christos G. Tsagas$^{2,1}$ and Georgios Fanaras$^{2}$\\ {\small $^1$DAMTP, Centre for Mathematical Sciences, University of Cambridge}\\ {\small Wilberforce Road, Cambridge CB3 0WA, UK}\\ {\small $^2$Section of Astrophysics, Astronomy and Mechanics, Department of Physics}\\ {\small Aristotle University of Thessaloniki, Thessaloniki 54124, Greece}}

\maketitle

\begin{abstract}
We consider Friedmann-like universes with torsion and take a step towards studying their stability. In so doing, we apply dynamical-system techniques to an autonomous system of differential equations, which monitors the evolution of these models via the associated density parameters. Assuming relatively weak torsion, we identify the system's equilibrium points. These are found to represent homogeneous and isotropic spacetimes with nonzero torsion that undergo accelerated expansion. We then examine the linear stability of the aforementioned fixed points. Our results indicate that Friedmann-like cosmologies with weak torsion are generally stable attractors, either asymptotically or in the Lyapunov sense. In addition, depending on the equation of state of the matter, the equilibrium states can also act as intermediate saddle points, marking a transition from a torsional to a torsion-free universe.
\end{abstract}

\section{Introduction}\label{sI}
Extensions of general relativity that go beyond the boundaries of the Riemannian geometry, by allowing for an asymmetric affine connection, have a long history in the literature. These studies introduce the possibility of spacetime torsion and its associated new degrees of freedom to the gravitational field (e.g.~see~\cite{BH} for a recent review). Therefore, it comes to no surprise that there are many applications of these theories to cosmology, in an effort to illuminate the role and the potential implications of torsion (as well as those of spin) for the evolution of the universe we live in. The topics of research interest, which have varied over the decades, range from the early universe and its initial singularity, to the large-scale kinematics and the late-time universal acceleration (see~\cite{G,P} for a representative though incomplete list).

Allowing for arbitrary torsion, introduces anisotropic degrees of freedom into the host spacetime. As a result, the spatially homogeneous and isotropic Friedmann-Robertson-Walker (FRW) cosmologies can only accommodate specific forms of torsion~\cite{T}. Vectorial torsion fields, determined by a single scalar function of time ($\phi=\phi(t)$ -- see \S~\ref{ssTF} here), are generally compatible with the FRW symmetries~\cite{KTBI}. The latter study focused on finding exact solutions for torsional Friedmann-like models. These were then combined with the primordial nucleosynthesis measurements to constrain the gravitational effects of torsion. Here, following on the work of ~\cite{KTBI}, we investigate the general qualitative behaviour of homogeneous and isotropic torsional cosmologies. Utilising the above named torsion scalar, $\phi=\phi(t)$, we parametrise the contribution of the torsion field to the universal expansion and to the total (effective) energy density of the universe. Then, assuming relatively weak torsion, we are able to recast the associated Einstein-Cartan equations into an autonomous dynamical system and identify its critical (fixed) equilibrium points. As expected, these include the familiar torsion-free Friedmann models, with varying 3-curvature and a nonzero cosmological constant. In the presence of (weak) torsion, on the other hand, we find that all the critical points correspond to spatially flat cosmologies and that they all undergo accelerated, de Sitter-like, expansion. Put another way, the torsional equilibrium states identified in this work are flat Friedmann-type universes, which are either $\Lambda$-dominated or filled with non-conventional matter (dark energy or phantom). This will change, however, if the weak-torsion assumption is relaxed (see \S~\ref{ssEPs} here).

Perturbing the aforementioned fixed points, we employ standard dynamical-system techniques to determine their linear stability and then complete the phase-space portraits of their dynamical evolution. Our results show that, with one exception that leads to a ``saddle point'', Friedmann-like cosmologies equipped with a weak torsion field are generally stable attractors, either asymptotically or in the Lyapunov sense. More specifically, the attractors correspond to accelerating universes with (weak) torsion, whereas the saddle point marks the transition from an accelerated torsional cosmology to a torsion-free (also accelerating) model.

The plan of the paper is as follows: In section 2 we introduce the underlying equations and define the torsion field in a Friedmann-type universe. The dimensionless variables, the associated autonomous dynamical systems and their critical points are defined and obtained in section 3. We study the stability of the critical points and identify some subtle issues surrounding equilibrium points with zero eigenvalues in section 4. There, we also provide the phase portraits of our dynamical study. Finally, we summarise our conclusions in section 5.

\section{Friedmann-like universes with torsion}\label{sF-LUT} 
The spatial homogeneity and isotropy of the Friedmann universes severely restricts the forms of torsion that they can accommodate naturally. In particular, the torsion fields allowed in an FRW cosmology must depend only on time and should have vanishing spacelike parts.

\subsection{The torsion field}\label{ssTF}
The general form of torsion permitted by the high symmetry of an FRW host has been given in~\cite{T}. Here, following~\cite{KTBI}, we will consider a sub-class of the allowed torsion fields, with the torsion tensor taking the form
\begin{equation}
S_{abc}= 2\phi h_{a[b}u_{c]}\,,  \label{Sabc}
\end{equation}
which falls into the class of the so-called vectorial torsion fields~\cite{CCSV}. Then, the associated torsion vector is given by
\begin{equation}
S_a= S^b{}_{ab}= -3\phi u_a\,.  \label{Sa}
\end{equation}
In the above, $\phi=\phi(t)$ is a scalar function of time,\footnote{The torsion scalar can in principle take positive or negative values, with the sign of $\phi$ determining the orientation of the torsion vector relative to the $u_a$-field (see Eq.~(\ref{Sa}) above as well as~\cite{KTBI}).} $u_a$ is a timelike 4-velocity vector (i.e.~$u_au^a=-1$) and $h_{ab}=g_{ab}+u_au_b$ is the symmetric spacelike tensor orthogonal to it (i.e.~$h_{ab}=h_{ba}$, $h_{ab}u^b=0$ and $h_a{}^a=3$). We also note that an immediate consequence of (\ref{Sabc}) and (\ref{Sa}) is that $S_a$ becomes the sole carrier of the torsion effects.

The common assumption is that spacetime torsion is induced by the spin of the matter, just like curvature is generated by the matter's energy-density contribution. Then, in an FRW-type cosmology, the Cartan field equations recast relations (\ref{Sabc}) and (\ref{Sa}) into the expressions $\kappa s_{abc}=8\phi h_{c[a}u_{b]}$ and $\kappa s_a=12\phi u_a$ for the spin tensor ($s_{abc}=s_{[ab]c}$) and the spin vector ($s_a=s^b{}_{ab}$) respectively. These, in turn lead to the following close relations
\begin{equation}
S_{abc}- -{1\over4}\,\kappa s_{cba} \hspace{10mm} {\rm and} \hspace{10mm} S_a= -{1\over4}\,\kappa s_a\,,  \label{Ss}
\end{equation}
between torsion and spin in a Friedmann-like cosmology~\cite{KTBI}. Therefore, the two fields are directly proportional and they are both fully determined by the scalar function $\phi=\phi(t)$. With these in mind, we will explicitly focus on torsion.

\subsection{The $\Omega$-parameters}\label{ssOmPs}
In the presence of torsion, the analogues of the Friedmann and the Raychaudhuri equations, in a spacetime with nonzero spatial curvature and non-vanishing cosmological constant (i.e.~when $K,\Lambda\neq0$) take the form (see~\cite{KTBI} for details)
\begin{equation}
\left({\frac{\dot{a}}{a}}\right)^{2}= {\frac{1}{3}}\,\kappa\rho- \frac{K}{a^{2}}+ {\frac{1}{3}}\,\Lambda- 4\phi^{2}- 4\left({\frac{\dot{a}}{a}}\right)\phi  \label{Fried1}
\end{equation}
and
\begin{equation}
{\frac{\ddot{a}}{a}}= -{\frac{1}{6}}\,\kappa\left(\rho+3p\right)+ {\frac{1}{3}}\,\Lambda- 2\dot{\phi}- 2\left({\frac{\dot{a}}{a}}\right)\phi\,,  \label{Ray1}
\end{equation}
respectively. Therefore, torsion can affect the evolution of the FRW-like host in a variety of ways, depending on the sign and the magnitudes of the $\phi$ and $\dot{\phi}$.

Keeping in mind that $H=\dot{a}/a$ defines the (purely Riemannian) Hubble parameter, relation (\ref{Fried1}) recasts into the constraint
\begin{equation}
1= \Omega_{\rho}+ \Omega_{K}+ \Omega_{\Lambda}+ \Omega_{\phi}\,,  \label{Fried2}
\end{equation}
where $\Omega_{\rho}=\kappa\rho/3H^2$, $\Omega_K=-K/a^2H^2$ and $\Omega_{\Lambda}=\Lambda/3H^2$ are the familiar $\Omega$-parameters associated with the matter, the 3-curvature and the cosmological constant. In an analogous way, the dimensionless parameter
\begin{equation}
\Omega_{\phi}= -4\left(1+{\phi\over H}\right){\phi\over H}\,,  \label{Omegaphi}
\end{equation}
monitors the torsion contribution to the total (effective) energy density of our model. Then, $\Omega_{\phi}=0$ when $\phi=0$ (trivial case), or when $\phi/H=-1$. Also note that the above can be written as $\Omega_{\phi}=-4(1+\chi)\chi$, with the dimensionless variable $\chi=\phi/H$ measuring the contribution of the torsion field relative to that of the Hubble expansion. Finally, following (\ref{Omegaphi}), the torsion contribution to the total effective energy density of the host spacetime can be either positive or negative, depending on the sign of $\phi$ (among others).

\subsection{The deceleration parameter}\label{ssDP}
Given that $q=-\ddot{a}a/\dot{a}^2=-[1+(\dot{H}/H^2)]$ defines the (purely Riemannian) deceleration parameter, Raychaudhuri's formula (see Eq.~(\ref{Ray1}) in \S~\ref{ssOmPs}) leads to
\begin{equation}
qH^2= {1\over6}\,\kappa(\rho+3p)- {1\over3}\,\Lambda+ 2\dot{\phi}+ 2H\phi\,.  \label{Ray2}
\end{equation}
As with the Friedmann-like equations earlier, the overall effect of torsion on the deceleration/acceleration of the host spacetime depends on the sign and the magnitudes of $\phi$ and $\dot{\phi}$. Alternatively, one may combine expressions (\ref{Fried2}) and (\ref{Ray2}), together with the definitions of the $\Omega$-parameters given in \S~\ref{ssOmPs}, to arrive at
\begin{equation}
q= {1\over2}\,(1+3w)(1-\Omega_K)- {3\over2}\,(1+w)\Omega_{\Lambda}- {1\over2}\,(2+3w)\Omega_{\phi}- 2\left(1-{\dot{\phi}\over\phi^2}\right)\chi^2\,,  \label{tq1}
\end{equation}
where $w=p/\rho$ is the barotropic index of the matter. We also remind the reader that $\chi=\phi/H$ is the dimensionless ratio that monitors the strength of the torsion field relative to the Hubble expansion (see definition (\ref{Omegaphi}) in \S~\ref{ssOmPs} previously). As expected, in the absence of torsion, both of the above reduce to their familiar FRW counterparts~\cite{UL}.

\section{The autonomous system}\label{sAS}
Starting from the Friedmann equations given in \S~\ref{ssOmPs} earlier, one can arrive to an autonomous system of dynamical equations describing the phase-space evolution of the FRW-like models in terms of the four $\Omega$-parameters defined in the same section.

\subsection{The dynamical equations}\label{ssDEs}
Proceeding along the lines of~\cite{UL}, we will combine the torsional analogues of the Friedmann equations seen in \S~\ref{ssOmPs} with the conservation law of the matter density to obtain a set of dynamical equations for the $\Omega$-parameters defined there. To begin with, in the presence of torsion, the continuity equation of an FRW-like cosmology reads (see~\cite{KTBI} for the derivation)
\begin{equation}
{\dot{\rho}\over\rho}= -3(1+w)H- 2(1+3w)H\chi\,,  \label{tFRWcont}
\end{equation}
Using the above evolution formula, while bearing in mind that $\dot{H}/H= -(1+q)H$, the time-derivative of $\Omega_{\rho}$ reads
\begin{equation}
\dot{\Omega}_{\rho}= -\left[3(1+w)-2(1+q)+2(1+3w)\chi\right]H \Omega_{\rho}\,.  \label{dotOm1}
\end{equation}
In an analogous manner we obtain
\begin{equation}
\dot{\Omega}_K= 2qH\Omega_K\,, \hspace{20mm} \dot{\Omega}_{\Lambda}= 2(1+q)H\Omega_{\Lambda}  \label{dotOm23}
\end{equation}
and
\begin{equation}
\dot{\Omega}_{\phi}= 2\left({\dot{\phi}\over\phi}- \dot{H\over H}\right)\left(\Omega_{\phi}+2\chi\right)\,.  \label{dotOm4}
\end{equation}
Note that in deriving the latter of these formulae we have also used the auxiliary relation $\dot{\chi}/\chi=\dot{\phi}/\phi- \dot{H}/H$. Finally, combining expressions (\ref{dotOm1})-(\ref{dotOm4}) with Eqs.~(\ref{Fried2}) and (\ref{tq1}), it is fairly straightforward to show that $\dot{\Omega}_{\rho}+\dot{\Omega}_K+\dot{\Omega}_{\Lambda}+ \dot{\Omega}_{\phi}=0$, as expected.

\subsection{The case of weak torsion}\label{ssCWT}
We can simplify the system (\ref{dotOm1})-(\ref{dotOm4}) by assuming a weak torsion field, characterised by $|\chi|=|\phi|/H\ll1$ and $|\dot{\phi}/\phi|\ll |\dot{H}|/H$. In other words, we constrain both the torsion field and its rate of change. Then, definition (\ref{Omegaphi}) reduces to the linear expression $\Omega_{\phi}\simeq -4\chi$, with the latter taking positive or negative values depending on the sign of $\chi=\phi/H$ and with $|\Omega_{\phi}|\ll1$ always. In such a case, the set of (\ref{dotOm1})-(\ref{dotOm4}) reduces to
\begin{eqnarray}
\dot{\Omega}_{\rho}\simeq -\left[3(1+w) -2(1+q)\right]H\Omega_{\rho}\,, &\hspace{20mm}&  \dot{\Omega}_K= 2qH\Omega_K\,,  \label{dynsys1a}\\ \dot{\Omega}_{\Lambda}= 2(1+q)H\Omega_{\Lambda} \hspace{10mm} &{\rm and}& \hspace{10mm} \dot{\Omega}_{\phi}\simeq (1+q)H\Omega_{\phi}\,.  \label{dynsys2}
\end{eqnarray}
Our last step is to introduce the dimensionless ``time-variable'' $\eta=\ln(a/a_0)$ -- with ${\rm d}\eta=H{\rm d}t$~\cite{UL}, which recasts the above into the autonomous system
\begin{eqnarray}
\Omega^{\prime}_{\rho}\simeq -\left[3(1+w) -2(1+q)\right]\Omega_{\rho}\,, &\hspace{20mm}&  \Omega^{\prime}_K= 2q\Omega_K\,,  \label{autsys1a}\\ \Omega^{\prime}_{\Lambda}= 2(1+q)\Omega_{\Lambda} \hspace{10mm} &{\rm and}& \hspace{10mm} \Omega^{\prime}_{\phi}\simeq (1+q)\Omega_{\phi}\,,  \label{autsys1}
\end{eqnarray}
respectively. Here, primes indicate differentiation with respect to $\eta$. Finally, given that $\Omega_{\rho}$ can be always obtained algebraically from  Eq.~(\ref{Fried2}), one simply has to solve the system
\begin{equation}
\Omega^{\prime}_K= 2q\Omega_K\,, \hspace{10mm} \Omega^{\prime}_{\Lambda}= 2(1+q)\Omega_{\Lambda} \hspace{10mm} {\rm and} \hspace{10mm} \Omega^{\prime}_{\phi}\simeq (1+q)\Omega_{\phi}\,,  \label{autsys2}
\end{equation}
where
\begin{equation}
q\simeq {1\over2}\,(1+3w)(1-\Omega_K)- {3\over2}\,(1+w)\Omega_{\Lambda}- {1\over2}\,(2+3w)\Omega_{\phi}\,,  \label{tq2}
\end{equation}
at our adopted level of approximation (i.e.~having dropped the $\chi^2$-order term -- see Eq.~(\ref{tq1})). According to expression (\ref{tq2}), the torsional analogue of the familiar \textit{Einstein-de Sitter universe} (with $\Omega_K=0= \Omega_{\Lambda}$ and $w=0$) has $q=1/2-\Omega_{\phi}\simeq1/2$, due to the weakness of the torsion field. On the other hand, in the presence of (weak) torsion, the \textit{coasting universe} solution (with $K=0=\Lambda$ and $w=-1/3$) has $q\simeq -\Omega_{\phi}/2\neq0$.

\subsection{The equilibrium points}\label{ssEPs}
The equilibrium (fixed) points of the autonomous system (\ref{autsys1}) are solutions of the set $\Omega^{\prime}_K=0=\Omega^{\prime}_{\Lambda}= \Omega^{\prime}_{\phi}$, which recasts as
\begin{equation}
q\Omega_K= 0\,, \hspace{10mm} (1+q)\Omega_{\Lambda}= 0 \hspace{5mm} {\rm and} \hspace{5mm} (1+q)\Omega_{\phi}\simeq 0\,.  \label{autsys2}
\end{equation}
At the same time, the deceleration parameter is still given by (\ref{tq2}). With these in hand, may distinguish between the following three main alternatives:\\

\noindent \textbf{(i)} The ``trivial'' equilibrium configurations, namely $(\Omega_K,\Omega_{\Lambda},\Omega_{\phi})=(0,0,0)$ with $\Omega_{\rho}=1$ and $q=(1+3w)/2$, $(\Omega_{\rho}, \Omega_{\Lambda},\Omega_{\phi})=(0,0,0)$ with $\Omega_K=1$ and $q=0$ and $(\Omega_{\rho},\Omega_K,\Omega_{\phi})=(0,0,0)$ with $\Omega_{\Lambda}=1$ and $q=-1$. The first fixed point corresponds to the familiar (torsionless) \textit{Friedmann universes} with conventional matter, Euclidean spatial geometry and no cosmological constant. The second and the third are the classical \textit{Milne} and \textit{de Sitter} solutions respectively.\\

\noindent \textbf{(ii)} Assuming non-zero torsion, we demand that $\Omega_{\phi}\neq0$. This ensures that $q\simeq-1\neq0$ always (see Eq.~(\ref{autsys2}c)), which in turn implies that $\Omega_K=0$ at all times (see (\ref{autsys2}a)).\footnote{The constraints $q\simeq-1$ and $\Omega_K=0$ are a direct consequence of our weak torsion assumption. This is reflected in Eq.~(\ref{dotOm4}), which ensures that the aforementioned conditions do not apply for a general torsion field. In that case, however, one needs an evolution equation for the the torsion scalar ($\phi$), in order to proceed.} On the other hand, expression (\ref{autsys2}b) allows for $\Omega_{\Lambda}\neq0$ and therefore for a nonzero cosmological constant. On using the above, relations (\ref{Fried2}) and (\ref{tq2}) combine to give
\begin{equation}
\Omega_{\rho}\simeq -{1\over3(1+w)}\,\Omega_{\phi} \hspace{10mm} {\rm and} \hspace{10mm} \Omega_{\Lambda}\simeq 1- {2+3w\over3(1+w)}\,\Omega_{\phi}\,,  \label{tcon}
\end{equation}
with $w\neq-1$ by default.\footnote{Following (\ref{tq1}), the value $w=-1$ of the barotropic index is also incompatible with our assumption that $\Omega_{\phi}\neq0$, which meant that $q=-1$ and $\Omega_K=0$.} In the case of a radiative fluid with $w=1/3$, the above constraints become $\Omega_{\rho}\simeq-\Omega_{\phi}/4$ and $\Omega_{\Lambda}\simeq1-3\Omega_{\phi}/4$. For pressure-free matter, that is for $w=0$, expressions (\ref{tcon}a) and (\ref{tcon}b) translate into $\Omega_{\rho}\simeq-\Omega_{\phi}/3$ and $\Omega_{\Lambda}\simeq1-2\Omega_{\phi}/3$ respectively. In either of the aforementioned cases $\bar{\Omega}_{\phi}$ has to be negative to guarantee ``ghost''-free matter with $\bar{\Omega}_{\rho}$ positive. Also, given that $|\Omega_{\phi}|\simeq4|\chi|\ll1$, we are always dealing with a $\Lambda$-dominated, spatially flat FRW-like universes with small amounts of matter (in the form of radiation or dust respectively) and weak torsion. Put another way, Eqs.~(\ref{tcon}) describe \textit{de Sitter}-like universes, which is in agreement with the value of the their deceleration parameters (recall that $q\simeq-1$, when $\Omega_{\phi}\neq0$).\\

\noindent \textbf{(iii)} Allowing for torsion, but switching the cosmological constant off (i.e.~assuming that $\Omega_{\phi}\neq0$ and $\Omega_{\Lambda}\equiv0$), the fixed point defined by (\ref{tcon}) has $q\simeq-1$, $\Omega_K=0$,
\begin{equation}
\Omega_{\rho}= -{1\over2+3w} \hspace{10mm} {\rm and} \hspace{10mm} \Omega_{\phi}= {3(1+w)\over2+3w}\,,  \label{tFP3}
\end{equation}
with $w\neq-1,-2/3$. Demanding that $\Omega_{\rho}>0$ always, namely excluding ghost-like matter with $\rho<0$, imposes the constraint $w<-2/3$ on the barotropic index. Then, assuming that torsion is subdominant at all times, relation (\ref{tFP3}b) leads to the following two sub-cases: \textbf{($\alpha$)} When $-4/3<w<-1$, the associated equilibrium point has
\begin{equation}
q\simeq -1\,, \hspace{7mm} \Omega_K= 0= \Omega_{\Lambda}\,, \hspace{7mm} \Omega_{\rho}\simeq -{1\over2+3w} \hspace{6mm} {\rm and} \hspace{6mm} \Omega_{\phi}= {3(1+w)\over2+3w}\,,  \label{alpha}
\end{equation}
where $\Omega_{\phi}>0$ and $\Omega_{\rho}>\Omega_{\phi}$; \textbf{($\beta$)} For $-1<w<-2/3$, the equilibrium point is still (formally) monitored by the same set of relations, though now $\Omega_{\phi}<0$ (with $\Omega_{\rho}>|\Omega_{\phi}|$). Note that the above given two fixed points correspond to spatially flat FRW-like universes, with weak torsion and non-conventional matter. The latter has positive energy density, but negative pressure and negative total gravitational energy density (i.e.~$w<-1/3\Leftrightarrow\rho+3p<0$ in both cases). This explains the de Sitter-like expansion (with $q\simeq-1$) of the associated solutions, despite the absence of a cosmological constant. Finally, we should note that when $w<-1$, we are dealing with the so-called ``phantom'' matter~\cite{CKW}.

\section{Stability analysis}\label{sSA}
In dynamical terms, the fixed points identified in the last section may be stable attractors, unstable repulsors, or intermediate saddle points. We can determine the stability of the aforementioned equilibrium configurations by perturbing them and then studying their response.

\subsection{Perturbing the equilibrium points}\label{ssPEPs}
To begin with, let us go back to the set of (\ref{autsys2}) and (\ref{tq2}). Substituting the latter into each one of Eqs.~(\ref{autsys2}a)-(\ref{autsys2}c), we obtain
\begin{equation}
\Omega_K^{\prime}\simeq (1+3w)\Omega_K- (1+3w)\Omega_K^2- 3(1+w)\Omega_K\Omega_{\Lambda}- (2+3w)\Omega_K\Omega_{\phi}\,,  \label{F1}
\end{equation}
\begin{equation}
\Omega_{\Lambda}^{\prime}\simeq 3(1+w)\Omega_{\Lambda}- 3(1+w)\Omega_{\Lambda}^2- (1+3w)\Omega_{\Lambda}\Omega_K- (2+3w)\Omega_{\Lambda}\Omega_{\phi}  \label{F2}
\end{equation}
and
\begin{equation}
\Omega_{\phi}^{\prime}\simeq {3\over2}\,(1+w)\Omega_{\phi}- {1\over2}\,(2+3w)\Omega_{\phi}^2- {1\over2}\,(1+3w)\Omega_{\phi}\Omega_K- {3\over2}\,(1+w)\Omega_{\phi}\Omega_{\Lambda}\,,  \label{F3}
\end{equation}
respectively. The next step is to introduce perturbations around the fixed-point solutions ($\bar{\Omega}_K,\bar{\Omega}_{\Lambda}, \bar{\Omega}_{\phi}$) obtained in \S~\ref{ssEPs}. More specifically, we set
\begin{equation}
\Omega_K= \bar{\Omega}_K+ \omega_K\,, \hspace{10mm} \Omega_{\Lambda}= \bar{\Omega}_{\Lambda}+ \omega_{\Lambda} \hspace{10mm} {\rm and} \hspace{10mm} \Omega_{\phi}= \bar{\Omega}_{\phi}+ \omega_{\phi}\,, \label{pertbOmega}
\end{equation}
with the quantities $\omega_K$, $\omega_{\Lambda}$ and $\omega_{\phi}$ representing homogeneous deviations from the equilibrium states.\footnote{Hereafter, overbars will always indicate variables evaluated at the equilibrium points.} Inserting the above into (\ref{F1})-(\ref{F3}), while taking into account that $\bar{\Omega}_K$, $\bar{\Omega}_{\Lambda}$ and $\bar{\Omega}_{\phi}$ satisfy a set formally identical to Eqs.~(\ref{F1})-(\ref{F3}), leads to the nonlinear propagation formulae
\begin{eqnarray}
\omega_K^{\prime}&\simeq& \left[1+3w-2(1+3w)\bar{\Omega}_K -3(1+w)\bar{\Omega}_{\Lambda}-(2+3w)\bar{\Omega}_{\phi}\right]\omega_K- 3(1+w)\bar{\Omega}_K\omega_{\Lambda} \nonumber\\  &&-(2+3w)\bar{\Omega}_K\omega_{\phi}- (1+3w)\omega_K^2- 3(1+w)\omega_K\omega_{\Lambda}- (2+3w)\omega_K\omega_{\phi}\,,  \label{nlomega'1a}
\end{eqnarray}
\begin{eqnarray}
\omega_{\Lambda}^{\prime}&\simeq& -(1+3w)\bar{\Omega}_{\Lambda}\omega_K+ \left[3(1+w)-6(1+w)\bar{\Omega}_{\Lambda} -(1+3w)\bar{\Omega}_K-(2+3w)\bar{\Omega}_{\phi}\right]\omega_{\Lambda} \nonumber\\ &&-(2+3w)\bar{\Omega}_{\Lambda}\omega_{\phi}- 3(1+w)\omega_{\Lambda}^2- (1+3w)\omega_{\Lambda}\omega_K- (2+3w)\omega_{\Lambda}\omega_{\phi}  \label{nlomega'2a}
\end{eqnarray}
and
\begin{eqnarray}
\omega_{\phi}^{\prime}&\simeq& -{1\over2}\,(1+3w)\bar{\Omega}_{\phi}\omega_K -{3\over2}\,(1+w)\bar{\Omega}_{\phi}\omega_{\Lambda} \nonumber\\ &&+\left[{3\over2}\,(1+w)-(2+3w)\bar{\Omega}_{\phi} -{1\over2}\,(1+3w)\bar{\Omega}_K -{3\over2}\,(1+w)\bar{\Omega}_{\Lambda}\right]\omega_{\phi} \nonumber\\ &&-{1\over2}\,(2+3w)\omega_{\phi}^2- {1\over2}\,(1+3w)\omega_{\phi}\omega_K- {3\over2}(1+w)\omega_{\phi}\omega_{\Lambda}\,,  \label{nlomega'3a}
\end{eqnarray}
of the perturbations themselves. The last three terms in each one of the above expressions are quadratic in $\omega$. Therefore, when the perturbations are relatively small (i.e.~for $|\omega|\ll|\Omega|$), the system (\ref{nlomega'1a})-(\ref{nlomega'3a}) linearises to
\begin{eqnarray}
\omega_K^{\prime}&\simeq& \left[1+3w-2(1+3w)\bar{\Omega}_K -3(1+w)\bar{\Omega}_{\Lambda}-(2+3w)\bar{\Omega}_{\phi}\right]\omega_K- 3(1+w)\bar{\Omega}_K\omega_{\Lambda} \nonumber\\ &&-(2+3w)\bar{\Omega}_K\omega_{\phi}\,,  \label{lomega'1a}
\end{eqnarray}
\begin{eqnarray}
\omega_{\Lambda}^{\prime}&\simeq& -(1+3w)\bar{\Omega}_{\Lambda}\omega_K+ \left[3(1+w)-6(1+w)\bar{\Omega}_{\Lambda} -(1+3w)\bar{\Omega}_K-(2+3w)\bar{\Omega}_{\phi}\right]\omega_{\Lambda} \nonumber\\ &&-(2+3w)\bar{\Omega}_{\Lambda}\omega_{\phi}  \label{lomega'2a}
\end{eqnarray}
and
\begin{eqnarray}
\omega_{\phi}^{\prime}&\simeq& -{1\over2}\,(1+3w)\bar{\Omega}_{\phi}\omega_K -{3\over2}\,(1+w)\bar{\Omega}_{\phi}\omega_{\Lambda} \nonumber\\ &&+\left[{3\over2}\,(1+w)-(2+3w)\bar{\Omega}_{\phi} -{1\over2}\,(1+3w)\bar{\Omega}_K -{3\over2}\,(1+w)\bar{\Omega}_{\Lambda}\right]\omega_{\phi}\,,  \label{lomega'3a}
\end{eqnarray}
In what follows, we will use this set of differential equations to determine the (linear) stability of the fixed points identified in \S~\ref{ssEPs} earlier.

\subsection{Stability of fixed points with 
$\bar{\Omega}_{\phi}=0$}\label{ssSFP1}
Not surprising, adding (weak) torsion perturbations does not alter the standard stability behaviour of the spatially flat FRW universes. For instance, when $\bar{\Omega}_K=0= \bar{\Omega}_{\Lambda}=\bar{\Omega}_{\phi}$ --  see case~(i) in \S~\ref{ssEPs}, the linear system (\ref{lomega'1a})-(\ref{lomega'3a}) reduces to
\begin{equation}
\left(\begin{matrix}
\omega^{\prime}_K \\ \omega^{\prime}_{\Lambda} \\ \omega^{\prime}_{\phi}
\end{matrix}\right)\simeq
\left(\begin{matrix}
1+3w & 0 & 0 \\ 0 & 3(1+w) & 0 \\ 0 & 0 & {3\over2}\,(1+w)
\end{matrix}\right)
\left(\begin{matrix}
\omega_K \\ \omega_{\Lambda} \\ \omega_{\phi}
\end{matrix}\right)\,,  \label{EdS}
\end{equation}
when written in matrix form. Therefore, for conventional matter with positive gravitational energy density and $w>-1/3$ all three eigenvalues are positive, thus making the associated Friedmann solutions unstable equilibrium points (repulsors -- see point A in Fig.~\ref{fig:ph-sp1}). On the other hand, when $-1<w<-1/3$ the FRW universes are saddle points, while for $w<-1$ they become attractors. This is exactly what happens in the absence of torsion as well (e.g.~see~\cite{UL}).

\subsection{Stability of fixed points with 
$\bar{\Omega}_{\phi}\neq0$  and
$\bar{\Omega}_{\Lambda}\neq0$}\label{ssSFP2}
For our purposes, all the interesting scenarios have $\bar{\Omega}_{\phi}\neq0$. Let us therefore begin our stability investigation by assuming that $\bar{\Omega}_K=0$ and $\bar{\Omega}_{\Lambda}\simeq1-[(2+3w)\bar{\Omega}_{\phi}/3(1+w)]$ --  see case~(ii) in \S~\ref{ssEPs}. Then, the system (\ref{lomega'1a})-(\ref{lomega'3a}) recasts as
\begin{equation}
\omega^{\prime}_K\simeq -2\omega_K\,,  \label{lomega'1c}
\end{equation}
\begin{eqnarray}
\omega_{\Lambda}^{\prime}&\simeq& -(1+3w)\left[1-{2+3w\over3(1+w)}\,\bar{\Omega}_{\phi}\right]\omega_K- \left[3(1+w)-(2+3w)\bar{\Omega}_{\phi}\right]\omega_{\Lambda} \nonumber\\ &&-(2+3w)\left[1-{2+3w\over3(1+w)}\,\bar{\Omega}_{\phi}\right] \omega_{\phi}  \label{lomega'2c}
\end{eqnarray}
and
\begin{equation}
\omega_{\phi}^{\prime}\simeq -{1\over2}\,(1+3w)\bar{\Omega}_{\phi}\omega_K -{3\over2}\,(1+w)\bar{\Omega}_{\phi}\omega_{\Lambda}- {1\over2}\,(2+3w)\bar{\Omega}_{\phi}\omega_{\phi}\,. \label{lomega'3c}
\end{equation}
Recall that the above set describes linear perturbations around a fixed point that corresponds to spatially flat, $\Lambda$-dominated (i.e.~accelerating) FRW universe with a small amount of matter and (weak) torsion.

Suppose now that matter is highly relativistic radiation. In that case, $w=1/3$ and the linear system (\ref{lomega'1c})-(\ref{lomega'3c}) reads
\begin{equation}
\left(\begin{matrix}
\omega^{\prime}_K \\ \omega^{\prime}_{\Lambda} \\ \omega^{\prime}_{\phi}
\end{matrix}\right)\simeq
\left(\begin{matrix}
-2 & 0 & 0 \\ -2(1-{3\over4}\,\bar{\Omega}_{\phi}) & -4(1-{3\over4}\,\bar{\Omega}_{\phi}) & -3(1-{3\over4}\,\bar{\Omega}_{\phi}) \\
-\bar{\Omega}_{\phi} & -2\,\bar{\Omega}_{\phi} & -{3\over2}\,\bar{\Omega}_{\phi}
\end{matrix}\right)
\left(\begin{matrix}
\omega_K \\ \omega_{\Lambda} \\ \omega_{\phi}
\end{matrix}\right)\,,  \label{rad1a}
\end{equation}
in matrix form. Note that $\bar{\Omega}_{\phi}<0$ to ensure non-ghost matter with $\bar{\Omega}_{\rho}>0$ (see case (ii) in \S~\ref{ssEPs} earlier), while $|\bar{\Omega}_{\phi}|\simeq 4|\chi|$ always. Then, since $|\chi|= |\phi|/H\ll1$ (see \S~\ref{ssCWT} earlier), we may keep up to $\bar{\Omega}_{\phi}$-order terms in our analysis. Similarly, when dealing with a non-relativistic (pressure-free) fluid with $w=0$, the linearised system (\ref{lomega'1c})-(\ref{lomega'3c}) becomes
\begin{equation}
\left(\begin{matrix}
\omega^{\prime}_K \\ \omega^{\prime}_{\Lambda} \\ \omega^{\prime}_{\phi}
\end{matrix}\right)\simeq
\left(\begin{matrix}
-2 & 0 & 0 \\ -(1-{2\over3}\,\bar{\Omega}_{\phi}) & -3(1-{2\over3}\,\bar{\Omega}_{\phi}) & -2(1-{2\over3}\,\bar{\Omega}_{\phi}) \\
-{1\over2}\,\bar{\Omega}_{\phi} & -{3\over2}\,\bar{\Omega}_{\phi} & -\bar{\Omega}_{\phi}
\end{matrix}\right)
\left(\begin{matrix}
\omega_K \\ \omega_{\Lambda} \\ \omega_{\phi}
\end{matrix}\right)\,,  \label{dust1a}
\end{equation}
with $\bar{\Omega}_{\phi}<0$ and $|\bar{\Omega}_{\phi}|\ll1$. Next, we will use (\ref{rad1a}) and (\ref{dust1a}) to study the linear stability of their associated equilibrium points.

\subsubsection{The case of $\omega_K=0$}\label{sssCSF}
Current observations strongly favour a universe with nearly flat spatial sections. On these grounds, we may (temporarily) ignore the effects of 3-curvature in our equations. Setting $\Omega_K\equiv0\Leftrightarrow\bar{\Omega}_K=0=\omega_K$, the linear systems (\ref{rad1a}) and (\ref{dust1a}) reduce to
\begin{equation}
\left(\begin{matrix}
\omega^{\prime}_{\Lambda} \\ \omega^{\prime}_{\phi}
\end{matrix}\right)\simeq
\left(\begin{matrix}
-4(1-{3\over4}\,\bar{\Omega}_{\phi}) & -3(1-{3\over4}\,\bar{\Omega}_{\phi}) \\
-2\,\bar{\Omega}_{\phi} & -{3\over2}\,\bar{\Omega}_{\phi}
\end{matrix}\right)
\left(\begin{matrix}
\omega_{\Lambda} \\ \omega_{\phi}
\end{matrix}\right)\,,  \label{rad1b}
\end{equation}
and
\begin{equation}
\left(\begin{matrix}
\omega^{\prime}_{\Lambda} \\ \omega^{\prime}_{\phi}
\end{matrix}\right)\simeq
\left(\begin{matrix}
-3(1-{2\over3}\,\bar{\Omega}_{\phi}) & -2(1-{2\over3}\,\bar{\Omega}_{\phi}) \\
-{3\over2}\,\bar{\Omega}_{\phi} & -\bar{\Omega}_{\phi}
\end{matrix}\right)
\left(\begin{matrix}
\omega_{\Lambda} \\ \omega_{\phi}
\end{matrix}\right)\,,  \label{dust1b}
\end{equation}
for radiation and dust respectively. The characteristic polynomials are $\mathcal{P}_1(\lambda)= -\lambda(\lambda+4-3\bar{\Omega}_{\phi}/2)$ for radiative matter and $\mathcal{P}_2(\lambda)=-\lambda(\lambda+3-\bar{\Omega}_{\phi})$ for a pressureless medium. Therefore, in both cases one of the eigenvalues is zero. In particular, we have $\lambda_1=0$ and $\lambda_2=-4+3\Omega_{\phi}/2<0$ in the case of radiation, while for pressure-free matter the eigenvalues are $\lambda_1=0$ and $\lambda_2= -3+\bar{\Omega}_{\phi}<0$. It is also straightforward to show that the eigenvectors corresponding to the zero eigenvalue of the radiation era are multiples of $v_1=(3,-4)$, while those associated with $\lambda_2$ are multiples of $v_2= ((4-3\bar{\Omega}_{\phi})/2\bar{\Omega}_{\phi},1)$. In the case of dust, on the other hand, the respective  eigenvectors are multiples of $u_1=(2,-3)$ and of $u_2= (2(3-2\bar{\Omega}_{\phi})/3\bar{\Omega}_{\phi},1)$.

\begin{figure}[tbp]
\centering \vspace{-1cm} {{\includegraphics[height=6cm,width=7cm]{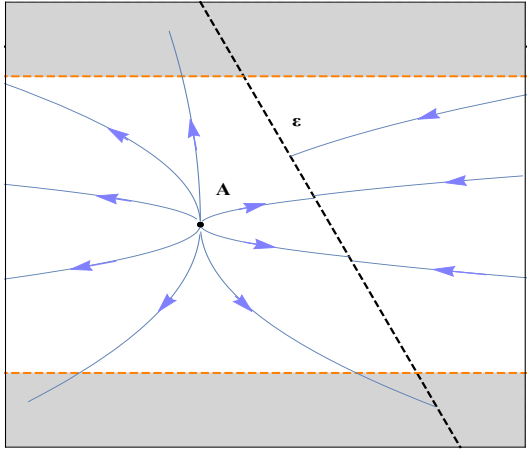}}} \caption{Phase-space diagram around two of the equilibrium points discussed in cases (i) and (ii) of \S~\ref{ssEPs}. The unstable fixed point A is a torsion-free, $\bar{\Omega}_{\rho}=1$ Friedmann universe with radiative matter or dust (e.g.~see~\cite{UL}, or \S~\ref{ssSFP1} here). The dashed diagonal line ($\varepsilon$) represents an accelerating, $\Lambda$-dominated, FRW-like cosmology with weak torsion (i.e.~$|\bar{\Omega}_{\phi}|\ll1$) and radiation (or dust), which is stable in the Lyapunoc sense. Note the shaded regions at the top and at the bottom of the diagram. These contain FRW-like universes with $|\bar{\Omega}_{\phi}|>1$ and therefore lie beyond the applicability range of this study (see \S~\ref{ssCWT}).}   \label{fig:ph-sp1}
\end{figure}

In the presence of a zero eigenvalue the stability of (\ref{rad1b}) and (\ref{dust1b}) is not straightforward to decide (see~\cite{BS} for an extensive discussion of the zero-eigenvalue problem). We will therefore attempt to obtain an answer by solving both of these systems analytically. Starting with the case of radiation, system (\ref{rad1b}) solves to give
\begin{equation}
\omega_{\Lambda}= {2\left(4-3\bar{\Omega}_{\phi}\right){\rm e}^{\alpha\eta}+ 3\bar{\Omega}_{\phi}\over8-3\bar{\Omega}_{\phi}}\, \mathcal{C}_1- {3\left(4-3\bar{\Omega}_{\phi}\right) \left(1-{\rm e}^{\alpha\eta}\right)\over2\left(8-3\bar{\Omega}_{\phi}\right)}\, \mathcal{C}_2 \label{rad1}
\end{equation}
and
\begin{equation}
\omega_{\phi}= -{4\bar{\Omega}_{\phi}\left(1-{\rm e}^{\alpha\eta}\right)\over8-3\bar{\Omega}_{\phi}}\,\mathcal{C}_1+ {8-3\bar{\Omega}_{\phi}\left(2-{\rm e}^{\alpha\eta}\right)\over8-3\bar{\Omega}_{\phi}}\,\mathcal{C}_2\,,  \label{rad2}
\end{equation}
where $\mathcal{C}_{1,2}$ are the integration constants and $\alpha=-(8-3\bar{\Omega}_{\phi})/2$. Based on our constraint that $|\bar{\Omega}_{\phi}|\ll1$, we deduce that $\alpha<0$ always. Consequently, at late times (i.e.~as $\eta\rightarrow\infty$), the above approaches the constant solution
\begin{equation}
\omega_{\Lambda}= {3\over8}\,\bar{\Omega}_{\phi}\,\mathcal{C}_1- {3\over4}\,\mathcal{C}_2 \hspace{10mm} {\rm and} \hspace{10mm} \omega_{\phi}= -{1\over2}\,\bar{\Omega}_{\phi}\,\mathcal{C}_1+ \mathcal{C}_2\,,  \label{ltrad1}
\end{equation}
which satisfies the condition $4\omega_{\Lambda}+3\omega_{\phi}=0$ as well. In other words, the perturbations $\omega_{\Lambda}$ and $\omega_{\phi}$ do not grow but instead tend to constant values. Moreover, solution (\ref{ltrad1}) resides within the sub-space of the eigenvector $v_1=(3,-4)$, which corresponds to the zero eigenvalue of (\ref{rad1b}). Consequently, solutions that start out near the equilibrium point remain close to it, although they never converge to the fixed point. Technically speaking, the system (\ref{rad1b}) is not asymptotically stable, but it is stable according to Lyapunov (see line ($\varepsilon$) in Fig.\ref{fig:ph-sp1}).

The same is also true for pressure-free matter. Indeed, when $|\bar{\Omega}_{\phi}|\ll1$, the late-time solution of system (\ref{dust1b}) reads
\begin{equation}
\omega_{\Lambda}= {1\over3}\,\bar{\Omega}_{\phi}\,\mathcal{C}_1- {2\over3}\,\mathcal{C}_2 \hspace{10mm} {\rm and} \hspace{10mm} \omega_{\phi}= -{1\over2}\,\bar{\Omega}_{\phi}\,\mathcal{C}_1+ \mathcal{C}_2\,,  \label{ltdust1}
\end{equation}
with $3\omega_{\Lambda}+2\omega_{\phi}=0$. As before, $\omega_{\Lambda}$ and $\omega_{\phi}$ are constants and the solution resides in the sub-space of the eigenvector $u_1=(2,-3)$, which in turn corresponds to the zero eigenvalue of the dust case (see matrix (\ref{dust1b}) above). Therefore, $\Lambda$-dominated Friedmann universes with dust, vanishing spatial curvature and torsion are also stable in the Lyapunov sense. In other words, solutions close to the aforementioned fixed point will remain in that vicinity and never diverge.

\subsubsection{The case of $\omega_K\neq0$}\label{sssCNSC}
Allowing for curvature perturbations and assuming that matter is highly relativistic radiation, the linear system is given by (\ref{rad1a}). Then, the characteristic polynomial is
$\mathcal{P}_1(\lambda)=-\lambda(\lambda+2) (\lambda+4-3\bar{\Omega}_{\phi}/2)$, with eigenvalues $\lambda_1=0$, $\lambda_2=-2$ and $\lambda_3=-4+3\bar{\Omega}_{\phi}/2<0$ (since $|\bar{\Omega}_{\phi}|\ll1$). When dealing with non-relativistic (pressure-free) matter, the linear system is (\ref{dust1a}) and the characteristic polynomial reads $\mathcal{P}_2(\lambda)= -\lambda(\lambda+2)(\lambda+3-\bar{\Omega}_{\phi})$. Here, the eigenvalues are $\lambda_1=0$, $\lambda_2=-2$ and $\lambda_3=-3+\bar{\Omega}_{\phi}<0$ (given that $|\bar{\Omega}_{\phi}|\ll1$).

As before, due to the zero eigenvalues, we will attempt to solve both linear systems analytically. Recalling that $|\bar{\Omega}_{\phi}|\ll1$, the late-time solutions (i.e.~as $\eta\rightarrow\infty$) read
\begin{equation}
\omega_K= 0\,, \hspace{10mm} \omega_{\Lambda}= {3\over8}\,\bar{\Omega}_{\phi}\,\mathcal{C}_1- {3\over4}\,\mathcal{C}_2 \hspace{5mm} {\rm and} \hspace{5mm} \omega_{\phi}= -{1\over2}\,\bar{\Omega}_{\phi}\,\mathcal{C}_1+ \mathcal{C}_2  \label{ltrad2}
\end{equation}
and
\begin{equation}
\omega_K= 0\,, \hspace{10mm} \omega_{\Lambda}= {1\over3}\,\bar{\Omega}_{\phi}\,\mathcal{C}_1- {2\over3}\,\mathcal{C}_2 \hspace{5mm} {\rm and} \hspace{5mm} \omega_{\phi}= -{1\over2}\,\bar{\Omega}_{\phi}\,\mathcal{C}_1+ \mathcal{C}_2\,,  \label{ltdust2}
\end{equation}
for radiation and dust respectively. Clearly, since $\omega_{\Lambda}$ and $\omega_{\phi}$ tend to finite constants and only $\omega_K\rightarrow0$, the above given solutions are also stable \`{a} la Lyapunov and not asymptotically stable. Moreover, the $3\times3$ solutions (\ref{ltrad2}) and (\ref{ltdust2}) are essentially supplementary to their $2\times2$ counterparts (given by (\ref{ltrad1}) and (\ref{ltdust1}) respectively), with $\omega_{\Lambda}$ and $\omega_{\phi}$ still residing in the subspace of the eigenvectors $v_1=(3,-4)$, and $u_1=(2,-3)$ respectively. Recall that the aforementioned eigenvectors correspond to the zero eigenvalues of $\mathcal{P}_1(\lambda)$ and $\mathcal{P}_2(\lambda)$ -- see \S~\ref{sssCSF} earlier.

Based of the results obtained in \S~\ref{sssCSF} and \S~\ref{sssCNSC}, we conclude that an accelerating, $\Lambda$-dominated FRW universe with a small amount of matter (in the form of radiation or dust) and a (weak) torsion field is stable in the Lyapunov sense, irrespective of whether its spatial-curvature perturbations are accounted for, or not.

\subsection{Stability of fixed points with 
$\bar{\Omega}_{\phi}\neq0$  and
$\bar{\Omega}_{\Lambda}=0$}\label{ssSFP3}
\begin{figure}[tbp]
\centering \vspace{-1cm} \subfloat[]{{\includegraphics[width=5cm]{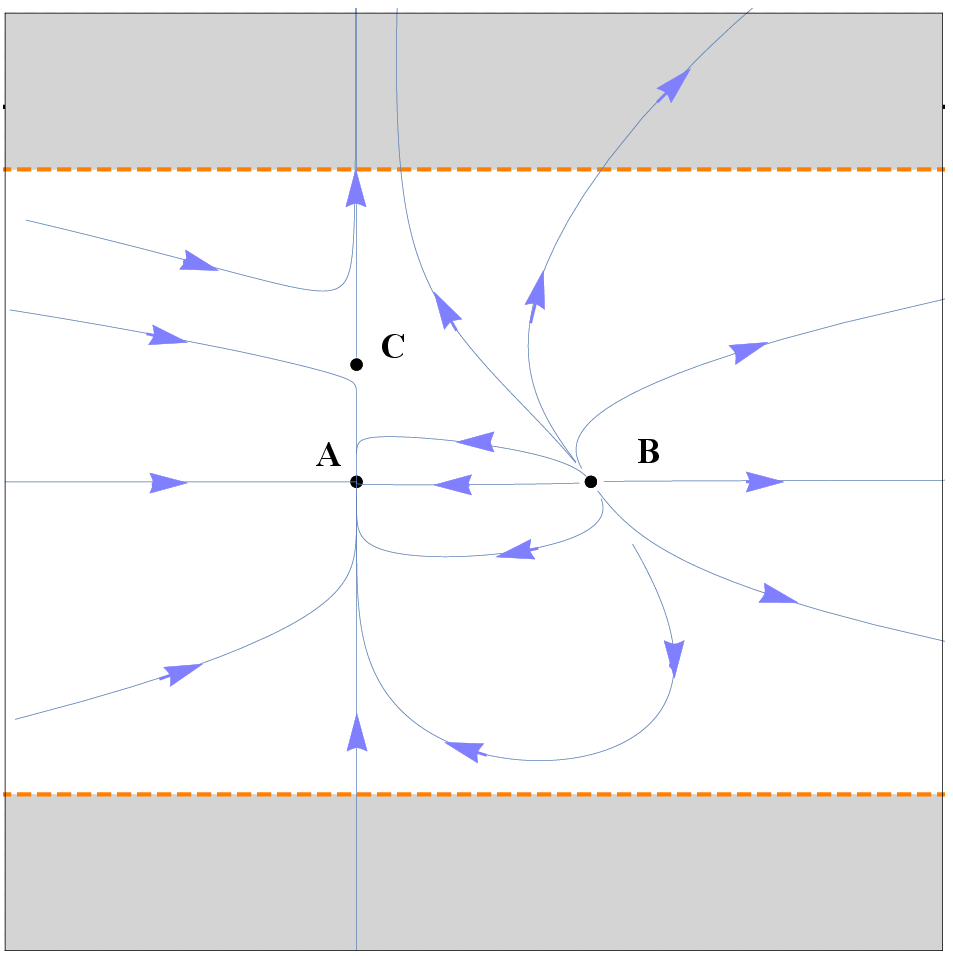}}}\qquad \qquad \subfloat[]{{\includegraphics[width=5cm]{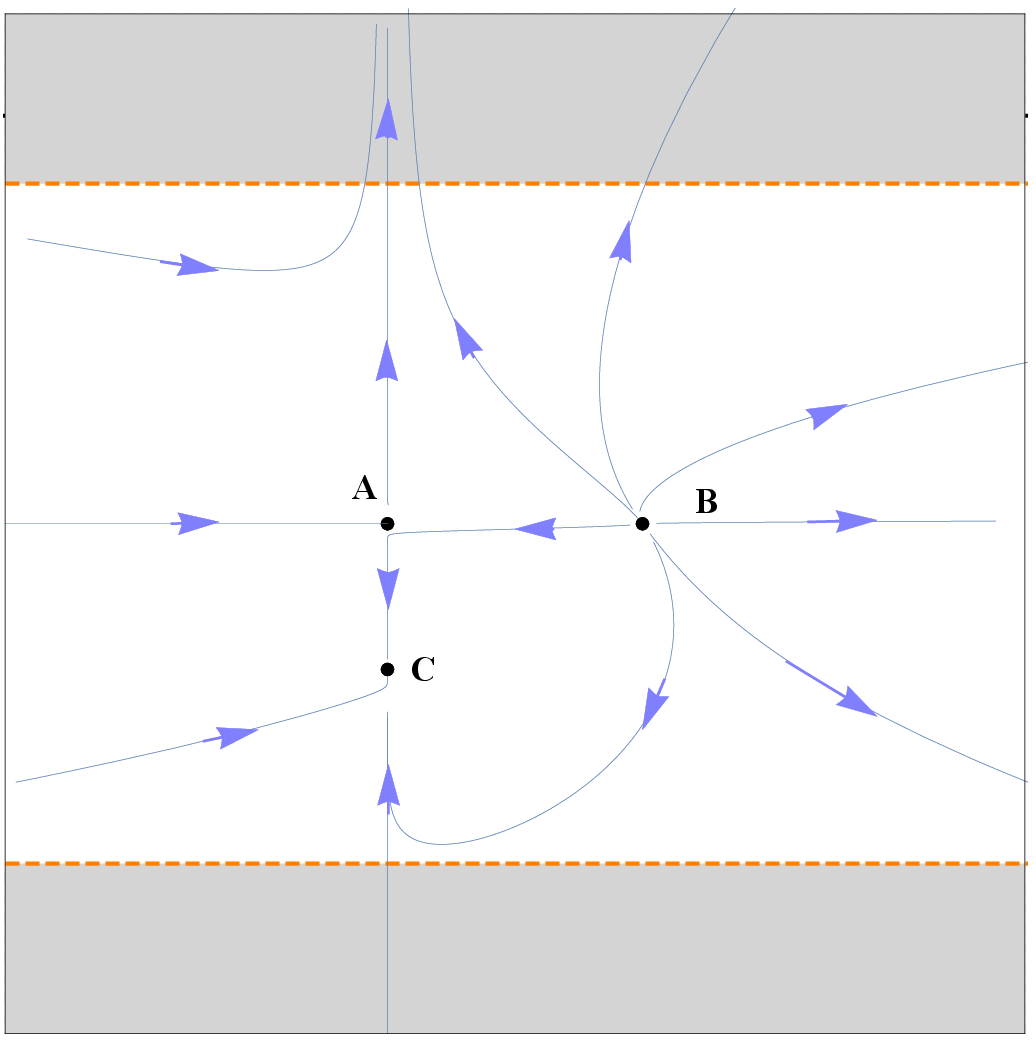}}} \caption{Stability diagrams of fixed points discussed in \S~\ref{ssEPs} (see cases (i) and (iii) there), with $-4/3<w<-1$ on the left-hand side and $-1<w<-2/3$ on the right. In Fig.~(a) the attractor A describes an accelerated, torsionless, flat FRW model filled with phantom matter (see \S~\ref{ssSFP1} here). Fixed point B is an unstable repulsor representing the Milne solution (e.g.~see~\cite{UL}), while C is an intermediate (transition) saddle point corresponding to an accelerating, spatially flat, FRW-like universe with weak torsion and phantom matter (see \S~\ref{ssSFP3}). In Fig.~(b), on the other hand, the saddle point A is a torsion-free Friedmann cosmology with zero 3-curvature and dark energy (see \S~\ref{ssSFP1} before), B is still the Milne repulsor and C is the attractor that represents a flat FRW-like model with weak torsion and dark energy. Note that the fixed point C has $\bar{\Omega}_{\phi}>0$ in Fig.~(a) and $\bar{\Omega}_{\phi}<0$ in Fig.~(b). Also, in both figures, the shaded regions contain torsional models with $|\bar{\Omega}_{\phi}|>1$, lying beyond the boundaries of our ``weak-torsion'' assumption.}   \label{fig:ph-sp23}
\end{figure}

Let us now switch the cosmological constant off, by setting $\bar{\Omega}_{\Lambda}=0=\omega_{\Lambda}$ at all times. Then, demanding that $\Omega_{\phi}\neq0$, $\Omega_{\rho}>0$ and $\Omega_{\rho}>|\Omega_{\phi}|$, we are in case (iii) of \S~\ref{ssEPs}. In other words, we are dealing with two families of equilibrium points having $-4/3<w<-1$ and $-1<w<-2/3$ (with $\Omega_{\phi}>0$ and $\Omega_{\phi}<0$ respectively). The former family corresponds to spatially Friedmann universes with torsion, dominated by ``phantom'' matter (since $w<-1$). In the latter case, on the other hand, the dominant matter component is of the dark-energy type (given that $-1<w<-2/3$). In either case, the associated cosmological models are accelerating with $q\simeq-1$. Also, dynamically speaking, both of the aforementioned sets of fixed points are monitored by the system (see Eqs.~(\ref{lomega'1a})-(\ref{lomega'3a}) in \S~\ref{ssPEPs})
\begin{equation}
\left(\begin{matrix}
\omega^{\prime}_K \\ \omega^{\prime}_{\phi}
\end{matrix}\right)\simeq
\left(\begin{matrix}
-2 & 0 \\ -{3(1+3w)\over2(2+3w)}\,(1+w) & -{3\over2}\,(1+w)
\end{matrix}\right)
\left(\begin{matrix}
\omega_{\Lambda} \\ \omega_{\phi}
\end{matrix}\right)\,,  \label{case3}
\end{equation}
with eigenvalues $\lambda_1=-2$ and $\lambda_2=-3(1+w)/2$. Consequently, the first family of equilibrium points (those with $-4/3<w<-1$) has one negative and one positive eigenvalue. This means that, in dynamical terms, phantom-dominated FRW-like  universes with zero spatial curvature and a weak torsion field are intermediate saddle points (see equilibrium state C in Fig.~(\ref{fig:ph-sp23}a)). On the other hand, the second family of fixed points (with $-1<w<-2/3$) has two negative eigenvalues. Therefore, flat Friedmann models with torsion and a dominant dark-energy component are stable attractors (see fixed point C in Fig.~(\ref{fig:ph-sp23}b)).\footnote{Keeping $\bar{\Omega}_{\Lambda}=0$, but allowing for nonzero perturbations (i.e.~assuming that $\omega_{\Lambda}\neq0$), we find that $\omega_{\Lambda}^{\prime}=0$ (see Eq.~(\ref{lomega'2a}) in \S~\ref{ssPEPs}). Incorporating this equation to (\ref{case3}) adds a zero eigenvalue to the system. This alters the nature of the stability, which is no longer asymptotic but of the Lyapunov type.}

\section{Discussion}\label{sD}
Dynamical system techniques have been extensively used to study the qualitative evolution of a wide range of cosmological solutions (e.g.~see~\cite{WE} for reviews). Among others, there have been applications to cosmologies with nonzero torsion (see~\cite{BBCCFT} and references therein.), Nevertheless, to the best of our knowledge, qualitative methods have not been used to study cosmological models based on pure Einstein-Cartan gravity, which is the simplest classical extension of general relativity. Here, we have attempted a step in this direction, by employing dynamical systems to study Friedmann-like universes with torsion. Before proceeding, however, one should bear in mind that the high symmetry of the FRW spacetimes imposes severe constraints on the form of the allowed torsion field~\cite{T}. Here, following on the work of~\cite{KTBI}, we have considered vectorial torsion, determined by a single scalar function of time ($\phi=\phi(t)$). We also introduced an effective density parameter ($\Omega_{\phi}$) to measure the torsion contribution to the total (effective) energy density of the universe. Then, assuming weak torsion (namely setting $\Omega_{\phi}\ll1$), we wrote the associated Einstein-Cartan equations as an autonomous system of differential equations.

The torsional equilibrium states of the aforementioned system corresponded to accelerating universes with zero spatial curvature. These cosmologies were either $\Lambda$-dominated, or they were filled with non-conventional matter, which satisfied a dark-energy/phantom equation of state. The stability analysis of these fixed points showed that they are all stable attractors, either asymptotically or \`{a} la Lyapunov, with the exception of the phantom-dominated solution. The latter was found to act as an intermediate saddle point, marking the transition from an accelerated torsional Friedmann-like universe to its (also accelerating) torsion-free counterpart.

Our results so far have been obtained under the assumption of weak torsion, with the dimensionless parameters $\phi/H$ and $\Omega_{\phi}=-4\phi/H$ restricted to values considerably smaller than unity. Relaxing these constraints, while remaining within the FRW framework, should add extra degrees of freedom to the solutions. The associated equilibrium points, in particular, are very likely to exhibit a richer and more versatile behaviour (like that reported in~\cite{KTBI} for example) and they should not necessarily identify themselves with accelerating, spatially flat spacetimes (see footnote~2 in \S~\ref{ssEPs}). However, in order to accommodate strong torsion fields to our analysis, one needs (among others) an evolution law for the torsion scalar. We will return to the investigation of the strong-torsion regime in our future work, as it goes beyond the scope of the present paper.\\

\textbf{Acknowledgements:} JDB was supported by the Science and Technology Facilities Council (STFC) of the UK. CGT wishes to acknowledge support from a visiting fellowship by Clare Hall and visitor support by DAMTP at the University of Cambridge, where the main part of this work was conducted.

\end{document}